# Machine Learning for Network Slicing Resource Management: A Comprehensive Survey

HAN Bin[1] and Hans D. Schotten[1,2]

(1. University of Kaiserslautern, 67663 Kaiserslautern, Germany;
2. German Research Center for Artificial Intelligence, 67663 Kaiserslautern, Germany)

**Abstract**
**The emerging technology of multi-tenancy network slicing is considered as an essential feature of 5G cellular networks. It provides network slices as a new type of public cloud services and therewith increases the service flexibility and enhances the network resource efficiency. Meanwhile, it raises new challenges of network resource management. A number of various methods have been proposed over the recent past years, in which machine learning and artificial intelligence techniques are widely deployed. In this article, we provide a survey to existing approaches of network slicing resource management, with a highlight on the roles played by machine learning in them.**

## 1 Introduction

As an emerging technology, network slicing is believed to be a key enabler and essential feature of the fifth generation (5G) cellular networks. Proposed by the Next Generation Mobile Networks (NGMN) Alliance as an end-to-end (E2E) concept, network slicing is involved across the radio access network (RAN) and the core network (CN). It refers to operating and maintaining multiple logically independent virtual telecommunication networks on the top of a shared physical infrastructure, in order to provide enhanced heterogeneity, flexibility, scalability, profitability and security of future network services. This requires both the network resources and network functions to be highly countable, divisible and isolatable, which can be realized by the modern network function virtualization technologies.

Since its first proposal, network slicing has triggered extensive research interest in various topics in the broad scope of wireless networking. This includes network architecture design, E2E slice orchestration and management, slice blueprint design, slice lifecycle management, RAN virtualization, network resource management, slice isolation, mobility management, and cyber-security in network slicing.

In this article, we focus on the problems of resource management in network slicing, attempting to address the most significant challenges in this area and provide a timely and comprehensive survey to the state of the art. Especially, we will show how the modern techniques of machine learning and artificial intelligence are applied to assist the resource management in sliced wireless networks.

The rest of the paper will be organized as follows. Section 2 reviews the concept of network slicing in more details, and introduces the emerging multi-tenancy business case where network slices are provided as a public cloud service. In Section 3, we discuss two problems of resource management in network slicing, namely the slice admission control and the cross-slice resource management, and indicate the technical challenges in them. Existing intelligent solutions are studied in Section 4 thereafter. In the end, Section 5 closes this article with our conclusions.

## 2 Network Slicing and Multi-Tenancy Networks

### 2.1 Sliced 5G Network: Heterogeneous Services and Heterogeneous Requirements

The concept of network slicing refers to creating and maintaining multiple independent logical networks, i.e. "network slices", on the top of a shared physical network infrastructure. Every instance of the network slice, according to the definition of NGMN [1], is defined by a set of network functions and the resources to run them. These network functions and resources form a complete instantiated logical network, to meet certain network characteristics required by the service instance(s), which is realized within or by the network slice. Different network slice instances can be, fully or partially, physically or logically, isolated from each other in the perspectives of control, traffic, resources, etc. Furthermore, each slice instance can be individually tailored to fulfill the requirements by its service instance(s).

The feature of individual slice specification in network slicing plays a critical role in future 5G networks, due to the high heterogeneity of different 5G service types, i.e. enhanced mobile broadband (eMBB), massive machine type communications (mMTC), and ultra-reliable and low-latency communications (URLLC) [2]. These services generally have different requirements for technical performance, each being extreme in a different aspect, e.g., throughput, access capacity, and latency, as shown in **Fig. 1**. This implies highly heterogeneous specifications of resources and network functions for different types of slices. Indeed, even for a certain type of 5G service, the resource requirement can also vary from one service instance to another. Aiming at fulfilling the requirements of heterogeneous service instances simultaneously, the classical one-size-fits-all architecture that has been deployed in legacy Long Term Evolution/Long Term Evolution-Advanced (LTE/LTE-A) networks exposes significant lacks of flexibility and scalability, which can lead to low resource efficiency and therewith an unaffordable resource cost. Network slicing, in this context, has become an essential enabler of 5G networks.

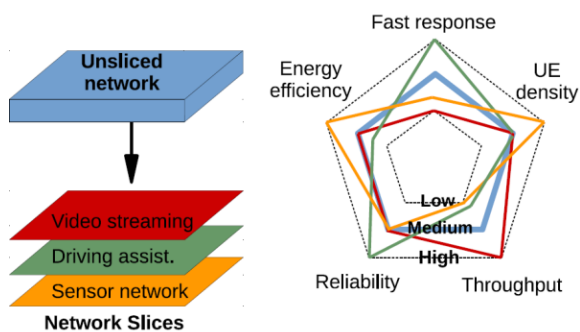

Figure 1. Network slicing enables heterogeneous and highly specialized services on top of a shared network infrastructure.

**2.2 Slice-as-a-Service: a New Public Cloud Environment**
In addition to the enhancement of resource efficiency, network slicing also makes it possible to decouple the provisions of wireless network infrastructure and network services. Instead of running and maintaining the network services by themselves, mobile network operators (MNOs) can lease network slices to multiple network slice tenants upon their requests. The tenants are therewith able to create network services and deliver them to the end customers without possessing their own network infrastructure, as illustrated in **Fig. 2**. The quality of service (QoS) of a leased slice is guaranteed by a service level agreement (SLA) between the MNO and the tenant, which defines the cost rate, the required minimal performances, and the penalty in case of SLA violation. This multi-tenancy network architecture introduces a new business mode that the network slices are provided as an emerging public cloud service, which is known as "slice-as-a-service" (SlaaS) [3].

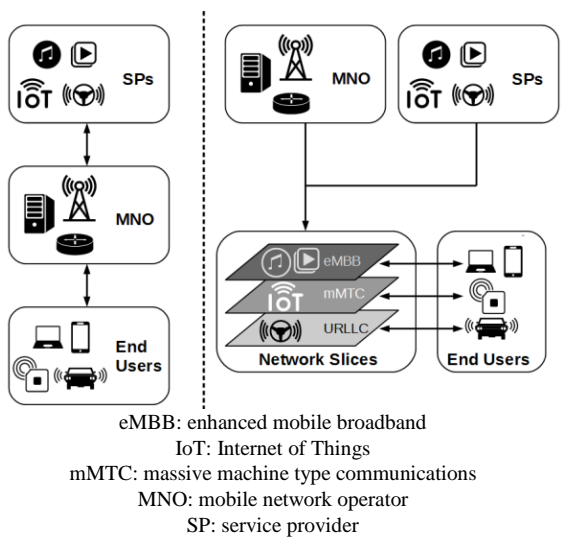

eMBB: enhanced mobile broadband
IoT: Internet of Things
mMTC: massive machine type communications
MNO: mobile network operator
SP: service provider

URLLC: ultra-reliable and low-latency communications
Figure 2. Traditional unsliced networks (left) and multi-tenancy sliced networks (right).

Despite of the similarity in many aspects to classical public cloud environments such as software-as-a-service (SaaS), platform-as-a-service (PaaS) and infrastructure-as-a-service (IaaS), SlaaS is distinguished from them in the complexity of resource management due to the heterogeneity of network slices, while the service instances in classical cloud environments are generally homogeneous. This challenges the efficient deployment of SlaaS and has triggered dense interest of research in recent years.

# 3 Resource Management in Network Slicing

**3.1 Classification of Approaches**
In an architectural perspective, efforts that have been made towards efficient resource management in sliced networks can be generally classified into two categories: the slice admission control and the cross-slice resource allocation (**Table 1**).

The former one consists of methods focusing on the issue that the limited resource pool of a MNO may be overloaded by an overwhelming amount of tenant requests for slices, whereby the MNO has to select some requests for acceptance while declining the others. It has been demonstrated that the policy of such selection, a.k.a. the slice admission strategy, has a dominant impact on the overall resource efficiency and utilization rate of sliced networks. Advanced methods are therefore proposed to find the best strategy, in order to optimize the long-term overall network performance statistically.

Approaches in the latter class, in contrast, concentrate on the active slices that have already been created and leased to tenants. The real-time traffic load of every individual slice is universally time varying, exhibiting stochastic dynamics. This phenomenon, known as the slice elasticity, enables the MNO to overbook slices to tenants for a diversity gain that improves the resource efficiency and overall revenue. To realize slice overbooking and jointly maximize the short-term performance of all active slices, it calls for methods that efficiently share network resources among slices in a real-time and dynamic fashion.

On the other hand, in perspective of the decision making mechanism, for both the admission control and cross-slice resource allocation, there are two types of approaches available: 1) policy-based decision and 2) auction-based decision (Table 1).

In policy-based approaches, the MNO provides a standard list of prices for slices (in case of admission control) or resources (in case of cross-slice resource allocation), which is consistent for all tenants, and the decision of admission/allocation is made according to the MNO's resource management policy under the current system state. In case of admission control, the system state information usually consists of the amount of idle resources, the set of current active slices, and the queuing status of awaiting requests. In case of cross-slice resource admission, on the other hand, such information usually refers to the resource pool size, and the set of current active slices along with their instantaneous resource demands and utility rates.

In auction-based approaches, the MNO does not provide universal prices, but only a list of available slices/resources. Instead, the tenants shall propose their own bids for their requested slices/resources. These bids are periodically collected and evaluated by the MNO, and the winner(s) of the auction will be granted the requested slice/resources. To guarantee a minimal revenue of operating the network infrastructure, lowest bids are universally required by the auction-based approaches.

Table 1. A summary of existing works on resource management in network slicing

| Reference | Slice admission control | Cross-slice resource allocation | Policy-based | Auction-based | Note |
|---|---|---|---|---|---|
| [4] | N | Y | Y | N | User admission control on every individual slice according to tenant-specific policies to allocate resources cross slices. |
| [5] | N | Y | Y | Y | Policy-based user admission control and user dropping on every slice to guarantee QoS; auction-based intra-slice resource allocation among users; budget-based inter-slice resource allocation. Dynamic cross-slice resource allocation not considered. |
| [6] | N | Y | Y | N | Grouping users according to behaviors and social relationships; bio-inspired methods to update the groups; policy-based cross-slice resource allocation according to group status |
| [7] | Y | N | Y | N | Uniformed slice size, binary slice admission control according to the active slice set, genetic algorithm to optimize the policy. |
| [8] | N | Y | Y | N | Deep Q-learning assisted allocation policy optimization. |
| [9] | Y | N | Y | N | A Markov model for policy-based slice admission control. |
| [10] | N | Y | Y | N | Jointly optimizing the base station bandwidth and the backhaul capacity as a bi-convex problem. |
| [11] | N | Y | N | Y | Non-cooperative auction among slices for network resources, implemented with OpenFlow |
| [12] | Y | N | Y | N | Q-learning assisted slice admission control policy optimization. |
| [13] | N | Y | Y | Y | A preliminary conference version of [5]. |
| [14] | N | Y | N | Y | A two-level slicing mechanism with 1) a price competition among network chunks to determine resource prices and 2) an auction mechanism to allocate resources among slices. |
| [15] | N | Y | N | Y | Optimizing the resource price function to maximize the total profit of slices / the net social welfare of network. |
| [16] | N | Y | Y | N | Empirical investigation on the diversity gain in SlaaS. |
| [17] | N | Y | Y | N | Sharing RAN resources among users according to both base station assignment and slice assignment. User admission control on every slice to shape traffic and guarantee the QoS. |
| [18] | N | Y | Y | N | Splitting the policy optimization problem into two sub-problems, one from the MNO's perspective to maximize the revenue and the other on (every) tenant's side to minimize the cost. A distributed optimization is therefore achieved through a game-fashion iteration of price updating. Both resource constraints and service fairness are taken account of. |
| [19] | N | Y | Y | N | Optimize the RAN resource allocation policy taking into account of the resource-partitioning problem. |
| [20] | Y | Y | Y | N | A two-layer framework merging slice admission control and cross-slice resource allocation. |
| [21] | N | Y | Y | N | Allocating users to subcarriers across different MVNOs to maximize the overall network profit, assuming the cost proportional to both power and bandwidth. |
| [22] | Y | N | Y | N | Multiple queues for different slice types, taking into account impatient behavior of tenants. |
| [23] | N | Y | Y | N | Dynamic resource allocation based on deep neural network assisted traffic prediction. Data-driven black-box optimization. |
| [24] | Y | Y | Y | N | Optimizing RAN resource allocation among slices and non-sliced network, where admissions to slice requests are controlled w.r.t. the demanded resource efficiency. |
| [25] | Y | N | Y | N | Modeling MNO's revenue under policy-based slice admission control, analyzing the construction of optimal policy. |
| [26] | Y | N | Y | N | Studying the rational behavior of impatient tenants in policy-based slice admission control with multiple queues. |

### 3.2 Key Challenges

A main and generic challenge for policy-based methods for network slicing resource management is the high computational complexity. On one hand, for both admission control and cross-slice resource allocation, the utility function is generally non-convex with regard to the MNO's policy, eliminating any analytical solution of the global optimum. On the other hand, numerical solvers are also challenged by the complexity of the problem. Policy-based admission control problems, no matter with or without queuing mechanism, are binary programming problems where the MNO's decision is always either "0" for decline or "1" for admission. The policy-based cross-slice resource allocation problems, in comparison, are integer programming problems, where the amount of resource allocated to an arbitrary slice is always integer times of some atomic resource block. Both the problems are known to be NP-hard, leading to an unaffordable computational effort to optimize the policy through exhaustive search.

In comparison to policy-based methods, auction-based approaches are proven effective to reduce the computational complexity significantly. However, it generally requires a careful design of the auction mechanism and strict regulations, in order to mitigate drawbacks and risks that intrinsically root in the procedure of auction itself, such as multi-round auction overhead, biased bidding, and cheating [27], [28].

Additionally, although slice overbooking and cross-slice resource allocation allow the MNO to benefit from the load-driven elasticity of network slices, they also lead to risk of overloading the shared resource pool when traffic peaks simultaneous occur across multiple slices. In this case, the MNO becomes incapable to deliver guaranteed QoS to all active slices and therefore have to violate some SLAs, which implies paying penalty to the involved tenants. Such a risk must be taken into account as part of the opportunity cost of maintaining slices. In an extreme case, the opportunity cost of accepting a request for new network slice instance may overwhelm the revenue generated by the corresponding slice, and therefore the greedy strategy fails in admission control.

On the other hand, being too conservative in admission control also leads to the MNO's loss, due to a two-fold reason. First, it naturally implies a low resource utilization rate and low revenue. Second, since the tenants' need for slices does not simply vanish, the declined requests will usually be either reissued later, or buffered in a queue for delayed admission. No matter which design is used, under a low admission rate, declined requests will stack to cause serious congestions, and therefore significantly raise the average delay between the issuing and the admission of a request. As we have indicated in [22], after being awaiting for too long time, tenants will eventually lose their patience and interest in the MNO's service. In a competitive SlaaS market, such situation can probably lead to permanent loss of tenants.

Aiming at an optimal balance between the resource feasibility and the admission rate, the MNO must have a deep understanding in tenant behavior. This includes the characteristics of both active slices (e.g., load dynamics, lifetime distribution, etc.) and tenant requests (e.g., arriving rate, impatience, etc.). This not only calls for accurate models, but also further raises the computational complexity.

## 4 Machine Learning and Artificial Intelligence Methods

### 4.1 Reinforcement Learning

Since policy-based network slicing resource management procedures are typically Markov decision processes (MDPs) where a policy maps every specific system "state" to a corresponding "action" and therewith generates a "reward". In network slicing resource management problems, the reward function is generally non-convex over a huge policy space, as proven in [9]. Therefore, in this field people commonly choose to rely on Reinforcement Learning (RL), which is known for its high efficiency and convenient implementation in solving Markov decision problems.

A pioneering attempt of deploying RL to optimize the network slicing policy was given by [12], where the authors have demonstrated that their Q-Learning solver can efficiently approximate the optimal slice admission policy that maximizes the MNO's revenue and significantly outperform the benchmark of random policies. In comparison to the value iteration method that guarantees to achieve the optimum, the Q-Learning method is capable to be executed in an online learning fashion with a much more reasonable computation cost, with only a tradeoff of slight reduction in the revenue. Furthermore, RL algorithms can be designed model-free by appropriately selecting the reward functions, which makes them much more robust against imperfect estimations of the slicing statistics, as also demonstrated in [12].

The authors of [20] attempted to apply RL for cross-slice resource allocation, which they called cross-slice congestion control. Aiming at this, they have proposed a framework where the real-time slice elasticity is realized upon requests of every existing slice for the grant of more resource and the MNO makes policy-based decisions with regard to both the current resource availability and the slice priorities. In this way, the cross-slice resource allocation task is accomplished by an admission-control-like mechanism, where a Q-Learning method is proven to bring a significant gain in slice elasticity.

Cross-slice resource allocation was achieved in a more straightforward manner in [8], where the authors defined an "action" of the system as a specific allocation of radio resource to all existing slices instead of a binary decision like in slice admission. This design simplifies the system design, but leads to a significantly huger policy space and a high non-linearity of the reward function about the action. To cope with this issue, the authors adopted deep neural networks, as we will introduce later in Section 4.2.

**4.2 Artificial Neural Networks**
As the most important part of modern artificial intelligence technologies, artificial neural networks (ANN) are known to be efficient in modeling non-linear systems. This can be used to enhance RL methods into deep reinforcement learning (DRL) methods, such the deep Q-Learning method reported in [8].

Another common application of ANN is the model estimation and prediction of complex non-linear processes. The authors of [23] have given a typical example of ANN-based prediction in the field of network slicing resource management. In this work, they stacked three layers of three-dimensional convolutional neural networks (3D-CNN) to compose an encoder, which is cascaded with a decoder implemented by multi-layer perceptrons (MLPs). This encoder-decoder structured cognitive network is proven capable to predict service capacity requirement in a data-driven fashion with high accuracy, which helps the slice orchestrator to make decisions in slice admission control and cross-slice resource allocation. In contrast, legacy methods are only able to predict the mean traffic.

**4.3 Evolutionary Algorithms**
There are various methods, which rely on statistical evolutions based on learning from the system feedbacks to random strategies. They are commonly referred to as evolutionary algorithms, which is an important category of machine learning techniques.

One example of evolutionary algorithms' application in cross-slice resource allocation is given by [6], where the social relationship between different users attached to multiple network slices are updated in a dynamic and evolutionary manner. Based on these social relationships, users are clustered into groups in such a way that all users in the same group have similar characteristics in service requirement. This process helps in degrading and simplifying the complex model of resource requirement in a large-size sliced network, and therefore supports to optimize the resource allocation strategy.

In context of slice admission control, on the other hand, we have shown in our previous work [7] the effectiveness of genetic algorithms (GAs). By encoding every slice admission policy into a chromosome, i.e. a binary sequence, and applying a classical GA on a population of randomly generated chromosomes, it will recursively generate new generations of chromosomes (policies) that statistically converge towards an optimum. Furthermore, by manually introducing (an arbitrary) benchmark policies into the first generation, this GA-based mechanism is guaranteed to outperform the benchmark. It also shows good robustness against dynamic environments.

**4.4 Distributed Learning**
While all the aforementioned cases generally invoke a centralized learning process, some efforts have been made to distribute the learning process over different participators in the network slicing process, i.e. the MNO and different tenants/slices, in order to reduce the computational complexity.

A typical example is [11], where a RL process is executed simultaneously at every bidder (slice) to recursively update its bid for network resources. This so-called Exponential Reinforcement Learning (XL) algorithm is proven to converge to the unique Nash equilibrium of the auction game.

Similarly, the authors of [18] decomposed the cross-slice resource allocation problem into a revenue-maximizing problem of the MNO and a cost-minimizing problem of every slice. This sets up a game where a distributed evolutionary algorithm converges to the equilibrium.

Another instance is provided by [21], which invokes the famous Binary Particle Swarm Optimization (BIPSO) algorithm, which allows to jointly update the resource assignments to different users in a distributed cross-learning manner, i.e. in each iteration, the resource assignment to every specific user will be updated according to the resource assignments to other users in the last iteration. Such iterative update continues until the utility requirement is satisfied. The authors have shown that the BIPSO is computationally efficient in solving the policy-based cross-slice radio resource allocation optimization problem.

# 5 Future Challenges

Beyond the successes that have already been made, there are still many open issues and potentials for further successes of machine learning in the field of network slicing resource management, as we will name some of them below.

**5.1 AI-Enhanced Optimization in More Complex Admission Control Scenario**
As it has been pointed out, complex features of slices/tenants, such as elasticity [20] and impatience [22], will lead to challenges in modeling their behavior, even under ideal assumptions such as Poisson arrivals of traffic/service requests. In realistic scenarios, the request arrivals and slice/resource release are usually non-Markovian. This calls for a deeper understanding in the system behavior and better policy optimizers, which shall be provided by a better integration of ANN with RL methods, like the authors of [23] have done.

**5.2 Cooperative Game with Distributed Learning**
While existing applications of distributed learning in this field generally consider non-cooperative games where the Nash equilibriums are achieved, there is a great potential to adopt the concept of cooperative game, where tenants/slices can learn to make decisions in an organized and cooperative way, in order to maximize the global social welfare instead of their own interests. In this way, a Pareto optimum can be expected instead of the Nash equilibrium.

# 6 Conclusions

In this survey, we have discussed the resource management problem in multi-tenancy network slicing, introduced different types of approaches in this field, and extensively reviewed the existing works. Especially, we have shown how the modern techniques of machine learning and artificial intelligence could be applied in this field, and have named some open issues for potential future work.

## Biographies

**HAN Bin** (binhan@eit.uni-kl.de) received his B.E. degree in 2009 from Shanghai Jiao Tong University, China and his M.Sc. degree in 2012 from Darmstadt University of Technology, Germany. In 2016 he was granted the Ph.D. degree in electrical and information engineering from Kalsruhe Institute of Technology, Germany. Since July 2016 he has been with Institute of Wireless Communication, University of Kaiserslautern, Germany. His research interests are in the broad area of wireless networks and signal processing. HAN Bin has been involved in multiple European Union H2020 research projects and has published over 30 research papers.

**Hans D. Schotten** received the Diplom and Ph.D. degrees in electrical engineering from RWTH Aachen University, Germany in 1990 and 1997, respectively. Since 2007, he has been Full Professor and Head of the Institute of Wireless Communication at the University of Kaiserslautern, Germany. Since 2012, he has been Scientific Director at the German Research Center for Artificial Intelligence heading the Intelligent Networks Department.